\documentclass{vgtc}
\ifpdf%                                % if we use pdflatex
  \pdfoutput=1\relax                   % create PDFs from pdfLaTeX
  \usepackage{graphicx}                % allow us to embed graphics files
  \DeclareGraphicsExtensions{.pdf,.png,.jpg,.jpeg} % for pdflatex we expect .pdf, .png, or .jpg files
\else%                                 % else we use pure latex
  \usepackage{graphicx}                % allow us to embed graphics files
  \DeclareGraphicsExtensions{.eps}     % for pure latex we expect eps files
\fi%

\graphicspath{{figures/}{pictures/}{images/}{./}} % where to search for the images

\usepackage{microtype}                 % use micro-typography (slightly more compact, better to read)
\PassOptionsToPackage{warn}{textcomp}  % to address font issues with \textrightarrow
\usepackage{textcomp}                  % use better special symbols
\usepackage{mathptmx}                  % use matching math font
\usepackage{times}                     % we use Times as the main font
\usepackage{cite}                      % needed to automatically sort the references
\usepackage{tabu}                      % only used for the table example
\usepackage{booktabs}   
\usepackage{booktabs}
\usepackage{soul}
\usepackage{color}
\usepackage{amssymb}
\usepackage{amsmath}

\usepackage{algorithm,algorithmic}
\usepackage{subfigure}
\onlineid{7244}

\vgtccategory{Research}

\vgtcinsertpkg

\newcommand{\name}{News Kaleidoscope}
\newcommand{\person}{Gary}
\title{News Kaleidoscope: Visual Investigation of Coverage Diversity in News Event Reporting}

\author{Aditi Mishra\thanks{e-mail: amishr45@asu.edu} %
\and Shashank Ginjpalli\thanks{e-mail:sginjpal@asu.edu} %
\and Chris Bryan\thanks{e-mail:cbryan16@asu.edu}}
\affiliation{\scriptsize Arizona State University}
\teaser{
  \centering
  \includegraphics[width=.9\linewidth]{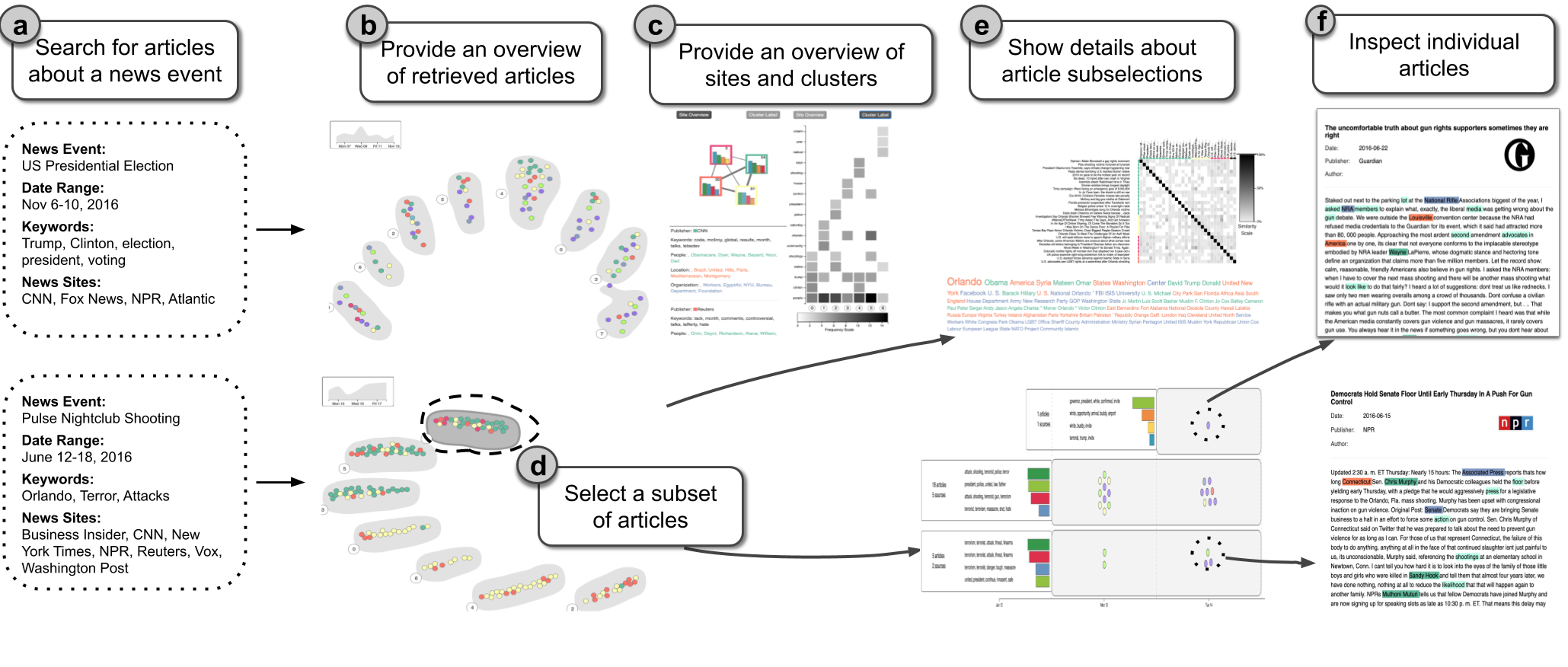}
  \caption{\name\ supports visual analysis of discrete news events with a focus on how coverage of an event varies. (a) After searching for news articles about an event of interest, (b) an overview ordination plot shows the retrieved articles. (c) A site level view and annotations for rendered clusters is also displayed. (d) Selecting a subset of articles (e) populates subselection views, enabling further investigation of coverage diversity. (f) Individual articles can also be inspected for detailed review.}
  \label{fig:teaser}
}
\abstract{
When a newsworthy event occurs, media articles that report on the event can vary widely---a concept known as \textit{coverage diversity}. To help investigate coverage diversity in event reporting, we develop a visual analytics system called \name{}. \name{} combines several backend language processing techniques with a coordinated visualization interface. Notably, \name{} is tailored for visualization non-experts, and adopts an analytic workflow based around subselection analysis, whereby second-level features of articles are extracted to provide a more detailed and nuanced analysis of coverage diversity.
To robustly evaluate \name, we conduct a trio of user studies. (1) A study with news experts assesses the insights promoted for our targeted journalism-savvy users.
 (2) A follow-up study with news novices assesses the overall system and the specific insights promoted for journalism-agnostic users. (3) Based on identified system limitations in these two studies, we refine \name{}’s design and conduct a third study to validate these improvements. Results indicate that, for both news novice and experts, \name{} supports an effective, task-driven workflow for analyzing the diversity of news coverage about events, though journalism expertise has a significant influence on the user’s insights and takeaways. Our insights developing and evaluating \name{} can aid future tools that combine visualization with natural language processing to analyze coverage diversity in news event reporting.
} % end of abstract

\CCScatlist{
  \CCScatTwelve{Human-centered computing}{Visu\-al\-iza\-tion}{Visu\-al\-iza\-tion techniques}{Treemaps};
  \CCScatTwelve{Human-centered computing}{Visu\-al\-iza\-tion}{Visualization design and evaluation methods}{}
}

\begin{document}


\maketitle
\section{Introduction}
In today's digital society, the ways that people access news is rapidly changing. Consumption on traditional platforms (print, radio, and television) is falling as consumption via websites and apps increases~~\cite{pew_state_of_the_media}. In addition to legacy news organizations adopting online presences, new digital-native ``born on the web'' publishers are utilizing social media platforms such as Facebook, Twitter, and YouTube for outreach, engagement, and sharing of stories~\cite{pew_state_of_the_media}.

A corollary to this increasing media pluralism is that, when a newsworthy event occur, articles reporting on the event will vary in different ways---a concept we refer to as \textit{coverage diversity}~\cite{humprecht2013more}. While coverage diversity can include structural aspects---e.g. the language a news article is written in and its length---it also includes thematic and framing aspects such as media bias. Systematic analyses have identified multiple types of media bias. For example, AllSides, a media watchdog group, classifies media bias into eleven categories, including the use of spin words, flawed logic, 
% unsubstantiated claims, slanted writing
sensationalism/emotionalism, and ad hominem attacks~\cite{allsides_eleven}. 

Our research aim is, given a news event of interest, to support the visual analysis the coverage diversity of news articles that report on it. Specifically, we consider coverage diversity in terms of the \textit{keywords} and \textit{entities } employed in reporting and their potential emotional biases, which are classified as emotionalism/sensationalism types of media bias~\cite{allsides_eleven}. We employ a design study methodology~\cite{sedlmair2012design} to develop and evaluate a novel visual analytics platform, called \name{},
% to interactively support querying for an event of interest to retrieve relevant news articles. 
% consisting of multiple coordinated visualizations, 
% including ordination, adjacency matrix, word cloud, emotion clusters, and annotated article text, 
which combines interactive visualizations and natural language processing (NLP) techniques
to analyze coverage diversity. 
% A backend supports event search and the application of natural language processing (NLP) techniques to retrieve news articles.

\name{} is designed based on a pre-study with journalism researchers. These users are interested in analyzing news reporting (and coverage diversity), but they are not visualization or NLP experts.
% to target users who are interested in news reporting (and cognizant of coverage diversity), but who are likely not visualization or NLP experts, such as journalism and political science researchers. 
% In contrast to general-purpose text and document visual analytics systems, \name{} is tailored for \textit{visualization non-expert users} who are interested in analyzing news articles, such as journalism and political science researchers.
% As such, we conduct a pre-study with intended users to motivate and constrain \name{}'s design. 
To identify ``polarities'' in coverage diversity, we adopt a workflow around selecting article subsets (\textit{subselections}), based on the premise that while articles about an event of interest will largely share top-level keywords and factoids, subselections will vary in second-level keywords, entities, and emotional stylings that enable a more nuanced understanding of coverage diversity.

To understand how \name{} supports the analysis of coverage diversity, we conduct an extensive set of evaluations.
We first evaluate \name{} with the journalism-savvy target users (Study~\#1), whom we call \textit{news experts}, to understand the specific types of insights promoted by the system. Then, to understand if and how \name{} generalizes to a broader user audience, we conduct a follow-up study with \textit{news novices} (Study~\#2). The participants in this study have little familiarity with U.S. news sites and therefore few preconceived assumptions about how coverage will differ when reporting on news events. 
Based on feedback from these studies, we implement a targeted set of design improvements to \name{} and conduct a validation study (Study~\#3) to verify their efficacy.

% To assess \name, we conduct a trio of qualitative user studies. (1) A usability study with \textit{news novices} assesses the overall system. Participants in this study have little familiarity with U.S. news sites and therefore few preconceived notions about potential biases attributed to them. (2) In a follow-up study, we conduct pair analytics sessions with \textit{news experts} (in  our case, journalism and political science professors) who are knowledgeable about perceived biases in American news sites. By holistically evaluating with these complementary user groups, we learn about how \name\ supports different insights based expertise, and identify a set of system drawbacks. We then implement a targeted set of system improvements, for example, by adding additional views to support summary and comparative analysis of news reporting, and then conduct (3) a validation study to verify our improvements. 
Our results indicate that \name\
effectively supports analysis of coverage diversity both for news experts and news novices, though interestingly it provides different benefits to each user group (e.g., for news experts, it supports validating hypotheses and assumptions about news sites).
% domain expertise enables more nuanced insights such as validating existing assumptions about news sites.
% with little background knowledge it is difficult to form strong or generalizable conclusions about news sites. For experts, existing background knowledge in journalism enables nuanced understanding of coverage diversity and can be used to validate existing assumptions about news sites. 
Based on our experiences in creating and evaluating \name{}, we discuss generalizable takeaways for visually analyzing coverage diversity in a news landscape that is increasingly online and diversified. In summary, the contributions paper include: (1)~we identify requirements for visually analyzing coverage diversity, which constitutes a current and real-world problem in journalism and media research, (2)~we design and implement a novel visualization system, \name{}, for analyzing news articles about an event of interest in the context of coverage diversity, (3)~based on a series of robust evaluations with both news experts and novices, we learn about design guidelines and implications for analyzing coverage diversity using visualization and NLP.
% of \name{}.

% \aditi{\color{blue}One of the reviewers said that we don't make it clear how is this system different from other text analysis systems.
% \begin{itemize}
    % \item We provide more comparative views. Other systems look into what the corpus is trying to say and finding dominant topics or issues over time or identifying relevant topics. Our aim isn't that, plus we also provide more comparative views. Site level similarity, article level similarity, heatmaps, graph etc. This comparison would not just help in doing what other text analysis systems do but also gives better coverage or viewpoints of similarities and differences.)
    % \item Most papers look into sentiment analysis, but we are trying to identify spin words or loaded words using emotion analysis. 
% \end{itemize}
% Also I am not sure where to mention it but this system as per the user study was found to be more useful for journalism researchers. Hence, this system is a very appropriate example of AI in journalism and can be sold that way.}

% \cite{sedlmair2012design} \hl{As per Sedlmair et al’s well-known 2012 Design Study Methodology paper, “a design study paper does not require a novel algorithm or technique contribution. Instead, a proposed visualization design is often a well justified combination of existing techniques.” Per that paper, we contribute the following: analysis, real-world problem, design, validation, and reflection. 
% }
\section{Related Work}
\label{sec:rw}

\textbf{Bias and Framing in News Reporting.}
% There is a long history of media bias in in American news reporting~\cite{kuypers2013partisan}.
% , however the proliferation of media sources in today's society---both online and via traditional mediums such as radio, television, and newspaper---has made it trivial for consumers to ``silo'' into media bubbles. In part because of this, a 2019 Pew Research Center survey indicated that one-fifth of both Democrats and Republicans only get political news from sources with like-minded audiences~\cite{gallup_2020_pathways}, increasing the potential for biased reporting.
Biased reporting can have substantial societal impacts. For example, a 2007 study by DellaVigna and Kaplan~\cite{dellavigna2007fox} found evidence that media bias 
% (specifically for right-leaning Fox News) 
had a significant effect on voting in the 1996 and 2000 presidential elections. More recently, a 2015 study~\cite{glynn2015} showed that exposure to biased news can lead to intolerance of dissent and group polarization.

Today, watchdog organizations such as AllSides~\cite{allsides} and the Center for Media and Public Affairs~\cite{cmpa} assess media bias and framing, primarily via qualitative assessments. In contrast, academic research has employed data-driven approaches to quantify bias in news reporting. A seminal 2005 paper by Groseclose and Milyo~\cite{groseclose2005measure} assigned bias scores to news organizations by counting think-tank citations in articles, scoring the majority of studied outlets as left-leaning. 
A recent survey paper~\cite{hamborg2019automated} reviews several automated approaches for identifying media bias in news articles, including ones that blend NLP with visualization, and 
% While we quantify bias in terms of emotionalism,
% (using a mature NLP library, see Section~\ref{sec::backend_server}),
recent work has investigated the use of machine learning for identifying and classifying bias (e.g., by analyzing the propaganda techniques used by news agencies~\cite{da-san-martino-etal-2020-prta}).
% according to 18 propaganda techniques used by news agencies.
% For example, An et al.~\cite{an2012visualizing} studied news site bias based on Twitter subscriptions and interaction patterns. 
% They visualized news sites along a political spectrum using co-subscription relationships inferred by Twitter links, finding similar results to Groseclose and Milyo. 
% As an example of news-focused visualization tools, LingoScope~\cite{diakopoulos2013visual} and Compare Clouds~\cite{diakopoulos2015compare} support the analysis of media by visualizing keywords distributions between two media groups or news sites.
% While we quantify bias in terms of emotionalism,
% (using a mature NLP library, see Section~\ref{sec::backend_server}),
% recent work~\cite{da-san-martino-etal-2020-prta} has proposed a machine learning model classified according to 18 propaganda techniques used by news agencies.
% ; it uses complex machine learning architecture to achieve identification of propaganda in news agencies. 
% While this research is currently not open-sourced, 
% \name{} supports integration of such emerging models to analyze different forms of bias.
% Future models can easily be integrated into systems like \name{} to support analysis of coverage diversity for different forms of bias.

\textbf{Visualizing News Reporting and Text Data.}
News visualization tools support the analysis of reporting. For example,  LingoScope~\cite{diakopoulos2013visual} and Compare Clouds~\cite{diakopoulos2015compare} enable the comparative analysis by visualizing keywords distributions between media groups or news sites. TimeMines~\cite{swan2000timemines} generates event timelines based on semantic features such as keywords and entities present in the text of articles (similarly, Chieu and Lee~\cite{chieu2004query} support time-based event extraction via keyword queries). While \name{} is a news visualization tool, its intended goal differs from these systems.

% While we quantify bias in terms of emotionalism,
% % (using a mature NLP library, see Section~\ref{sec::backend_server}),
% recent work~\cite{da-san-martino-etal-2020-prta} has proposed a machine learning model classified according to 18 propaganda techniques used by news agencies.

News-focused interfaces build upon the broader community of visualization of text data. Such visualizations normally leverage text mining and/or NLP techniques, including keyword extraction, entity recognition, event and topic modeling, sentiment analysis, and document similarity/clustering~\cite{kucher2018state, liu2018bridging, cao2016overview},
% , kucher2015text
to transform unstructured text corpora into derived text (meta)data that is suitable for visual analysis.
% For example, the characterization of emotional style we use is similar to what is used in the PEARL system~\cite{zhao2014pearl}, which visualizes a user's tweet history based on emotions.
% While \name{} opts to let users directly query for news articles using keyword-based search
% While \name\ omits a pipeline for identifying and extracting events/topics (instead opting to allow the user to directly search for articles about an event of interest), 
Several general-purpose visualization systems have been designed for high-level summarization and browsing of events/topics via aggregate visualizations~\cite{luo2010eventriver, dou2012leadline, cui2014hierarchical}.
% cui2011textflow, wei2010tiara, 
% Other techniques directly plot individual documents as discrete data points~\cite{cao2016overview}. Comparative document visualizations juxtapose a small set of documents (usually two), allowing the user to compare how they are alike and how they differ. 
% For example, literature fingerprinting~\cite{keim2007literature} visualizes computed feature values from a text document in a succinct visualization strip using a diverging color scale to highlight similarities and differences in the structure of two documents when they are arranged side-by-side. 
Alternatively, a collection of documents can be plotted as discrete data points using techniques such as matrices~\cite{alexander2014serendip}, clustering~\cite{stasko2008jigsaw}, 
% xu2003document
parallel coordinates~\cite{dou2011paralleltopics}, force-directed layouts~\cite{lee2012ivisclustering}, 
% alsakran2011streamit
and dimensionality reduction or ordination~\cite{chen2009exemplar, kim2016topiclens}. Several of these techniques require computing the pairwise similarities between documents---see Cao and Cui for an overview~\cite{cao2016overview}. In \name\ we employ a multi-weight aggregate distance metric that can be interactively adjusted.
% allow the user to interactively weight several distance metrics, resulting in a novel multi-weight distance metric to compute pairwise document similarities.

\textbf{Contextualizing \name\ to Previous Work.}
% For example, articles are generally written about a specific event of interest, allowing for targeted search/retrieval. 
% For a collection of retrieved articles, coverage diversity can be explored by analyzing subselections. 
% Specifically, \name{} supports analysis of subselections based on extracted keywords, entities, and emotions.
In contrast to prior-mentioned news and text visualization systems, which support tasks such as the broad summarization of events over time~\cite{luo2010eventriver} or support comparing the keywords used by two media sites~\cite{diakopoulos2013visual},
\name{}'s design is tailored towards a specific demographic and goal: people who are interested in analyzing the coverage diversity of news reporting, but who are likely not visualization or NLP experts. In contrast to general-purpose visualization interfaces for text and document corpus analysis, news articles have special semantics---for example, articles reporting on a news event likely share similar top-level keywords and emotions, but it may be necessary to investigate second-level nuances in the data to understand if and how coverage diversity is occurring. Further, we carefully design \name{} to balance its analytical capabilities with a user experience and visualization designs that are accessible to the target user base, by following a rigorous design study methodology~\cite{sedlmair2012design} that results in generalizable takeways demonstrating how the visual analysis of coverage diversity differs from general text visualization, particularly when accounting for the user expertise.

\section{Design Requirements Analysis}
\label{sec:ta}

To motivate a visual analytics design for investigating coverage diversity, we interviewed a trio of news experts: journalism professors who research or teach on news reporting and media (two experts are currently assistant professors, and the third is a full professor).

A significant problem in the journalism community (stated by each participant) is that there are a lack of computational tools or visualization software specifically designed for analyzing of reporting various styles or themes (i.e., coverage diversity) in news-based reporting. While the participants knew computational processes existed for analyzing text data (e.g., NLP algorithms), they had little familiarity with the technical aspects (leading to high uncertainty and limited trust). 
%A problem mentioned by each is the lack of computational tools to make sense of large text collections. Even journalism researchers largely have little familiarity or technical understanding of machine learning and NLP, and there is little presence in their domain way of advanced or customized visualization software to support analysis of media documents. For each professor, asked  about the way they search for articles and the various difficulties they face when exploring a large corpus of news documents in the context of coverage diversity. The common practice for our interviewees was to search for news articles of interest via keyword-based queries in search engines, before skimming or reading through the text of retrieved articles.
Based on these interviews, we derived a set of five design requirements (\textbf{DR}s) to support visual analysis specifically in the context of analyzing news reporting for coverage diversity. For each DR, we provide a brief justification about how it supports the intended domain (i.e., visualization non-experts who want to analyze news articles). Section~\ref{sec:hsd} describes the software stack in detail.

\textbf{DR1: Retrieve articles about an event of interest.}
The common ``first step'' for our participants in their analytic workflows was to search for news articles about a topic of interest via keyword-based queries in search engines, then skimming or reading through the text of retrieved articles. This workflow lies in contrast to some previous systems for document visualization, which begin by showing an aggregate view of the entire collection (e.g.,~\cite{luo2010eventriver, dou2012leadline}). Since such a summarization perspective is extraneous for the current domain (i.e., the participants already know what topic they want to search for, and what its high-level keywords are), we omit a summary view and instead \textit{allow users to directly query for a topic of interest based on high-level keywords and other constraints (date, source, etc.)}.

% To investigate a news event, the first step is retrieving articles that report on the event. All the professors search for events of their interest based on a keyword based query in search engines, thus they wanted \name{} to have a feature through which they could search for relevant news articles. Thus \textit{\name{} should account for the same since the news experts will be familiar with its high-level keywords, and can further tailor search results by adding or removing keywords as desired}.

%In \name, we implement search functionality via a sidebar control panel with keyword-based search and other filtering options (e.g., by media company). We intentionally omit ``event browsing'' visualization which would aggregate or cluster articles, similar to EventRiver~\cite{luo2010eventriver} and LeadLine~\cite{dou2012leadline}. While \name\ can easily be extended to integrate such functionality, such interfaces are more suited for browsing a collection of multiple events. In contrast, news experts are likely focusing on a single event, will be familiar with its high-level keywords, and can further tailor search results by adding or removing keywords as desired.

\textbf{DR2: Provide a high-level overview of coverage diversity.}
Participants normally reviewed the retrieved articles about a news topic in an iterative manner (e.g., as a list), which could quickly become overwhelming. 
The participants described that \textit{providing an overview of the retrieved articles about the news event of interest, in a way that emphasises the general (i.e., high-level) trends or groupings of the coverage diversity, would provide an initial sense of the event's coverage diversity.} Such a view would also function to drive subsequent analysis, via an overview-plus-details workflow~\cite{shneiderman1996eyes}.

\textbf{DR3: Select subsets of articles to analyze coverage diversity polarities.}
% While dimensionality reduction can provide a high-level overview of coverage diversity, detail views are necessary for more in-depth analysis~\cite{shneiderman1996eyes}.
In addition to providing high-level overview of an event's coverage diversity, participants described wanting to investigate specific polarities in reporting---subsets of articles containing interesting biases or semantics compared to the rest of the reporting articles. Analyzing the second-level keywords in these polarities, as well as the presence of polarity-specific emotional media biases, would provide nuanced understanding of the coverage diversity for a news event. Participants also desired that such analyses be able to commpare both within and across article subsets. In other words, \textit{selecting article subsets should enable detailed analysis of coverage diversity polarities via tailored visualizations}.

\textbf{DR4: Provide data-level explanations to ensure trust and verification.}
For users not familiar with complex NLP and/or data mining processes, it is important to provide trust and interpretability of models and algorithms to non-expert users~\cite{lipton2018mythos}. In our discussions about these topics, our participants were wary about simply trusting the outputs of algorithms or models they had little insight into. For example, in discussing the clustering of news articles, participants wanted to know why an articles might be binned a certain way. Thus, \textit{when advanced computational techniques or models are employed, mechanisms should be employed to promote trust and interpretability of model recommendations and decisions.}

% Also since this system is intended for users like journalism experts who might find complex visual interface overwhelming, \textit{the interface should support showcasing the backend algorithm transparency by showing why an article might be clustered or classified in a certain way or if the number of clusters in the system is optimal.}
% In \name{}, the ``raw text'' data of news articles is made available on demand. To help explain why, for example, an article might be clustered or classified in a certain way, extracted keywords and entities are annotated via highlighting in the text. 

\textbf{DR5: Visualization complexity should account for user expertise.}
Like the majority of journalism researchers, our participants were not experts in advanced visualization interfaces. This means that overly complex or esoteric visual designs could can lead to a failure in conveying information succinctly and easily. Instead, \textit{interface designs should strike a balance between providing sufficient analytic capabilities while also being approachable and intuitive to use.} \name{} is built with these considerations in mind, and validated in the user studies described in Section~\ref{sec:user_study}.

\section{The \name\ System Design}
\label{sec:hsd}

We now describe \name's system design.
The system is a full-stack application with three primary facets: (1)~a data preprocessing step, (2) a backend server for data storage, query, and NLP-based computation, and (3) a frontend interface for visualization and interaction. For examples of how \name{} can be used to analyze coverage diversity, see the use case in Section~\ref{sec:use_case} and the demo video include in the supplemental materials.
% As \name\ follows the Visual Information-Seeking Mantra, as we start by providing an overview of retrieved articles, wherein the user can select article subsets for detailed analysis using subsequent views. However, before describing the front-end interface and interaction features of \name, we first detail the backend data processing and server setup.

\subsection{Data Corpus and Preprocessing}
When describing \name, we use a large news article dataset titled \emph{All the News}~\cite{thompson}. This text corpus contains $143,000$ articles published on $14$ news sites from 2015--2017 and includes both the article text and related metadata (title, news site name, author, publication date, article URL, etc.). The news sites are Western media organizations (the only non-American site is The Guardian) across a spectrum of perceived liberal-to-conservative political biases. Figure~\ref{fig:study_1_prestudy_responses} includes a list of the $14$ news sites and labels their ``media bias ratings'' as scored by the AllSides organization~\cite{allsides}. By using this diverse corpus as a dataset, we ensure that reporting on news events will likely have a high degree of coverage diversity.

For each article in the corpus, we apply a trio of heuristics as preprocessing steps: keyword identification, named entity recognition, and determination of emotional style. \textbf{Keywords} from an article are extracted via the Gensim NLTK library \cite{Loper02nltk:the}.
% which is based on the TextRank algorithm~\cite{mihalcea2004textrank}, and stored as a vector. 
\textbf{Named entity recognition} follows a similar process. We use the Stanford NER library~\cite{finkel2005incorporating} to extract named persons, locations, and organizations in each article. These are stored to a set of three vectors for each article: one for names, one for locations, one for organizations. Vectors are extracted using the bag of words technique for each article retrieved, where the presence or an absence of a word is encoded as a 1 or a 0, respectively.
While keywords and entities might at first glance seem redundant, there are important semantic differences. Keywords give a broad sense of the topical content of an article, but the individual words are unclassified. In contrast, entities are specifically binned into classes, providing the user with an explicit set of relevant, descriptive proper nouns.

The third heuristic characterizes the \textbf{emotional style} of each article. 
% Our approach is similar (though not identical) to Zhao et al.'s process of estimating affective expression for Twitter data~\cite{zhao2014pearl}. 
We use Plutchik's discrete categorical model~\cite{plutchik2001nature} to classify emotional states into eight primary emotions, organized as the following pairs: anger-fear, anticipation-surprise, joy-sadness, and trust-disgust. The NRC lexicon~\cite{mohammad2010emotions} (which is based on Plutchik's model) extracts and classifies emotional words contained in an article. We then compute the frequency of words for each emotion in the article. Specifically, we construct a $1\times8$ vector $(e_1,e_2,....e_8)$. Each $e_i$ equals $n_i/N$, where $n_i$ is the number of words for that particular emotion, and $N$ is the total number of words in the article (after stop-word removal). This means each article has an emotional vector of length $=8$, where each $e_i$ represents how much of that particular emotion exists in the article's text. Using this multi-dimensional characterization enables a much more robust analysis compared to simpler positive/negative sentiment models (e.g., the popular Stanford CoreNLP library~\cite{manning2014stanford}).

% The third heuristic extracts the \textbf{emotional style} of each article. We use the model very similar to the PEARL system \cite{zhao2014pearl}. Any news article is always approached by the writer with an intention to bring out a particular feeling of the reader for an entity described. Hence to capture the emotional style of an article we use Plutchik's discrete categorical model \cite{plutchik2001nature}. This classifies emotional states into 4 pairs which equals to 8 classes namely (anger-fear, anticipation-surprise, joy-sadness, trust-disgust). We use the lexicon based approach to identify words in text documents and find the emotion frequencies out. The reason we chose a lexicon based approach is mostly because of the fact that we wanted the system to be able to support different kinds of text documents as well some which might be more noisy than the news data where a lexicon based approach would be a very good method to classify.

% To estimate the basic emotion from the news article, we calculate a $1*8$ vector $(e_1,e_2,....e_8)$ where $e_i$ is computed from $N_i/N$, where $N_i$ is the number of emotional words of a particular class of emotion and $N$ is the total number of words in a document after cleaning it using NLP techniques. 

\subsection{Backend Server}
\label{sec::backend_server}

The backend server is built using Node.js~\cite{tilkov2010node} and acts as the storage and service layer between the processed data and the frontend interface. 
News articles---along with extracted keywords, entities, and emotional style vectors---are stored to a SQL database.
% The server supports the following computational processes:

% \subsubsection{Retrieving Articles about a News Event}
Based on a user query---which can include constraints such as date, keywords, news sites to include/exclude, and the number of articles to retrieve---we retrieve articles from the database (\textbf{DR1}) and rank them using TF-IDF~\cite{ramos2003using}. Specifically, we consider the query as a single small document and calculate the TF-IDF for each article in the database that meets the date/site constraints, using this to sort the articles by relevance. The desired number of articles are then returned. 
For this returned collection of articles, we compute their aggregate pairwise distances for the overview visualization in the interface (\textbf{DR2}). This heuristic is a combination of three independent distance metrics: the preprocessed (1) keyword and (2) entity vectors, and also (3) the temporal similarity of articles based on their publication dates. 

\begin{figure*}[t]
  \centering
  \includegraphics[width=.92\textwidth]{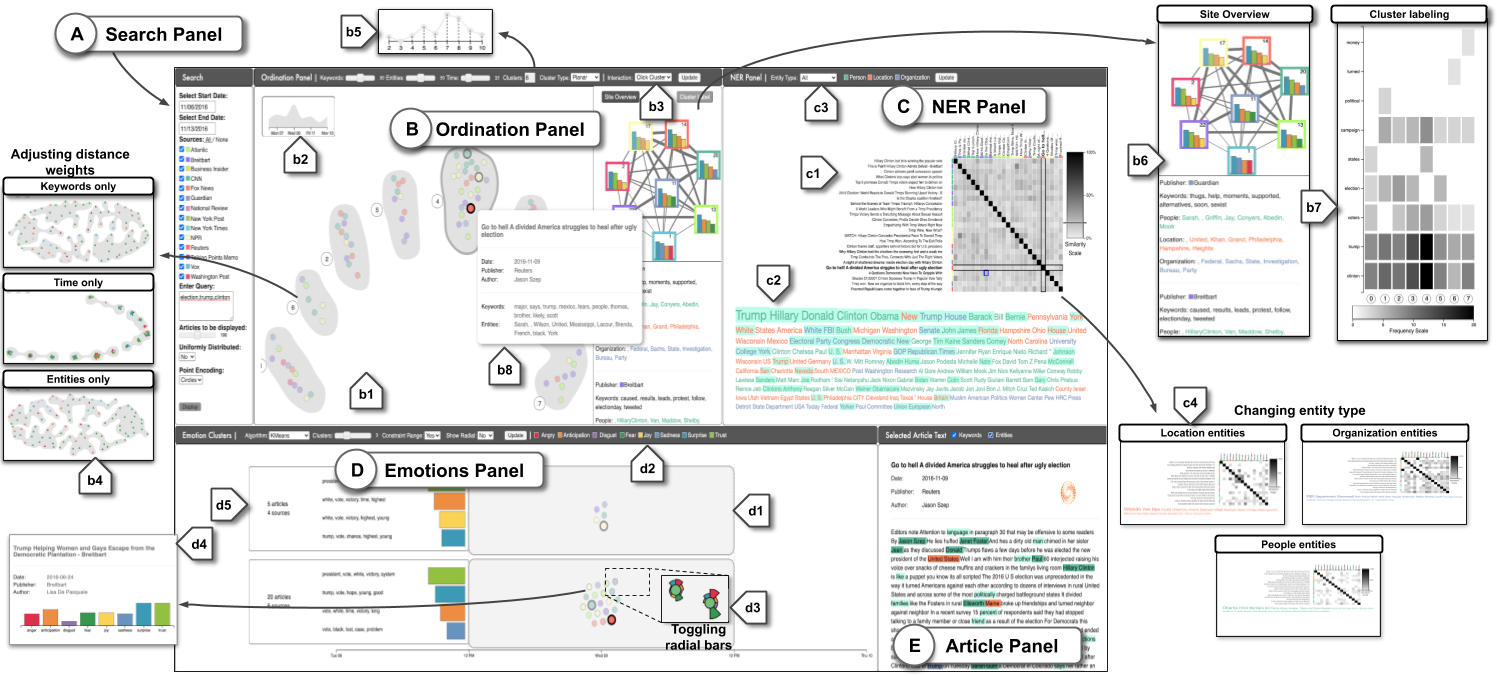}
  \caption{The \name{} interface consists of five panels to support \textsf{(A)} article search, \textsf{(B)} overview visualizations, \textsf{(C--D)} detailed analysis, and \textsf{(E)} inspection of individual articles.}  
  \label{fig:interface}
\end{figure*}

% \vspace{1.5mm} \noindent 
\textbf{Keyword Distance.} For all articles returned in a query, we compute the pairwise Jaccard similarity between their keyword vectors, normalized over the total number of words in each pair of articles. For articles $a_1$ and $a_2$, their keyword distance $d_k(a_1,a_2)$ represents how similar they are in terms of keywords on a scale between $[0,1]$.

% \vspace{1.5mm} \noindent 
\textbf{Entity Distance.} Similar to keyword distance, for each article pair we compute the normalized Jaccard similarity combined from each of three entity vectors (name, location, and organization), and then normalize the value over the total entities in the paired documents. Thus, the entity distance between articles $a_1$ and $a_2$ is given by $d_e(a_1,a_2)$, again on a scale from $[0,1]$.

% \vspace{1.5mm} \noindent 
\textbf{Temporal Distance.} Given that a news event generally happens over a discrete timeframe, we assume that reporting articles published in close temporal proximity (such as on the same day) might be more similar, as news coverage tends to evolve as follow-up investigation and reporting occurs~\cite{yang2009discovering}. Given a time range $R$ from which a set of articles are queried, we compute the temporal distance between articles $a_1$ and $a_2$ as $1 - \frac{date(a_1) - date(a_2)}{R}$. This distance $d_t(a_1,a_2)$ has a range between $[0,1]$.

% \vspace{1.5mm} \noindent 
\textbf{Aggregate Pairwise Distance.} For a set of retrieved articles, the pairwise keyword, entity, and temporal distances are independently computed and combined into a single aggregate distance $dist(a_1,a_2)$ which represents the overall similarity between two articles:
$$dist(a_1,a_2) = w_k\times d_k(a_1,a_2) + 
                    w_e\times d_e(a_1,a_2) +
                    w_t\times d_t(a_1,a_2)$$

Each distance metric is multiplied by a scaling weight ($w_k$, $w_e$, and $w_t$); these are interactively adjustable in \name{}'s frontend interface, enabling a multi-faceted exploration of articles based on desired user semantics. For example, setting $w_k=0$ would mean that the keyword distance would have no effect on the aggregate distance.
% In \name's frontend, users can interactively adjust each metric's weight, modifying how much it contributes to the aggregate distance. % This approach enables the multi-faceted exploration of articles based on multiple semantics.
One advantage of this aggregate multi-weight heuristic is that new similarity metrics can easily be added as desired. For example, a new metric for ``author similarity'' could be computed by analyzing a the historical style or content of a writer.

% \vspace{1.5mm} \noindent 
\textbf{Clustering Articles by Emotional Style.}
When article subselections are made (\textbf{DR3}), \name\ clusters article these articles by their emotional styles. To do this, the we calculate the pairwise similarity  in emotional style for articles, and then cluster the articles using \textit{k}-means clustering~\cite{macqueen1967some}.
% ~\cite{lloyd1982least}. 
The similarity is the Euclidean distance using the precomputed $1\times8$ emotional vector.

% This view offers a detailed view of the cluster or a set of articles the user has chosen by making a lasso. Each article has a similarity vector of dimension $1*8$ where 8 are the number of emotions we have accounted for. Assuming $n$ such articles were fetched, we apply a clustering algorithm, in our case the system supports both KMeans and Agglomerative custering for the users to discover patterns. The number of clusters is determined by the user by a slider which ranges from 2 to 5 clusters. The articles are then arranged according to force layout by their cluster number. \aditi{Where do we mention that the emotions we have are the words used by the author to invoke those emotions in the reader?} 

\subsection{Frontend Interface}
\label{sec:frontend_interface}

Figure~\ref{fig:interface} shows an overview of \name{}'s frontend interface. It consists of five linked panels \textsf{(A}--\textsf{E)}, designed to support the tasks \#1--6 described in Section~\ref{sec:ta}. As a note, this image shows the final design of \name{} that includes four additional features added based on participant feedback from Studies~\#1 and \#2---\textsf{(b5)}, \textsf{(b6)}, \textsf{(b7)}, and \textsf{(d5)}---which were  evaluated in Study~\#3. 

\textbf{\textsf{(A)} Search Panel.}
The search panel supports keyword-based searches for articles about news events (\textbf{DR1}). The user can set constraints/filters for date ranges, news sites, keywords, and the number of articles to retrieve. The user also has the option to return a uniform number of articles for each new sites: e.g., if $140$ articles are retrieved for $14$ sites, each site will return $10$ articles. Such a constraint provides a ``balanced'' distribution of article sources, but has the potential to lead to ``less relevant'' results if a news event was not extensively covered by a site.

\textbf{\textsf{(B)} Ordination Panel.}
The ordination panel provides an overview of retrieved articles (\textbf{DR2}). \textsf{(b1)} Article are encoded as circles, and colored by news site, and laid out via dimensionality reduction using aggregate pairwise distances. \name{} supports layout via multidimensional scaling (MDS)~\cite{kruskal1978multidimensional}, t-SNE~\cite{maaten2008visualizing}, and UMAP~\cite{mcinnes2018umap}. For the user studies, we employ MDS as it results in deterministic layouts, thus eliminating a potential confounding variable. Circles are clustered via \textit{k}-means clustering, with cluster hulls rendered using bubble sets~\cite{collins2009bubble}. We are motivated to employ this approach as it provides a scalable view of the retrieved articles, intuitively using proximity and groupings between data points to indicate their similarity. \textsf{(b2)} At upper left, an area chart shows the temporal distribution of retrieved articles.
 
\textsf{(b3)} At top, several widgets control display settings, including selecting the distance metric for cluster computation (either aggregate pairwise distance or x/y positioning), and updating the \textit{k} value to change the number of clusters. A trio of sliders adjust the scaling weights for the aggregate pairwise distances (the $w_k$, $w_e$, and $w_t$ values), which updates the layout in MDS plot \textsf{(b4)}. \textsf{(b5)} Hovering on the \textit{k} value input shows a tooltip that shows the silhouette score~\cite{rousseeuw1987silhouettes} for each value of $k$ between 2 and 10 (a higher score indicates better separation between clusters), as a way to explain to non-technical users what constitutes ``good'' choices for $k$.

\begin{figure*}[t]
  \centering
  \includegraphics[width=.9\textwidth]{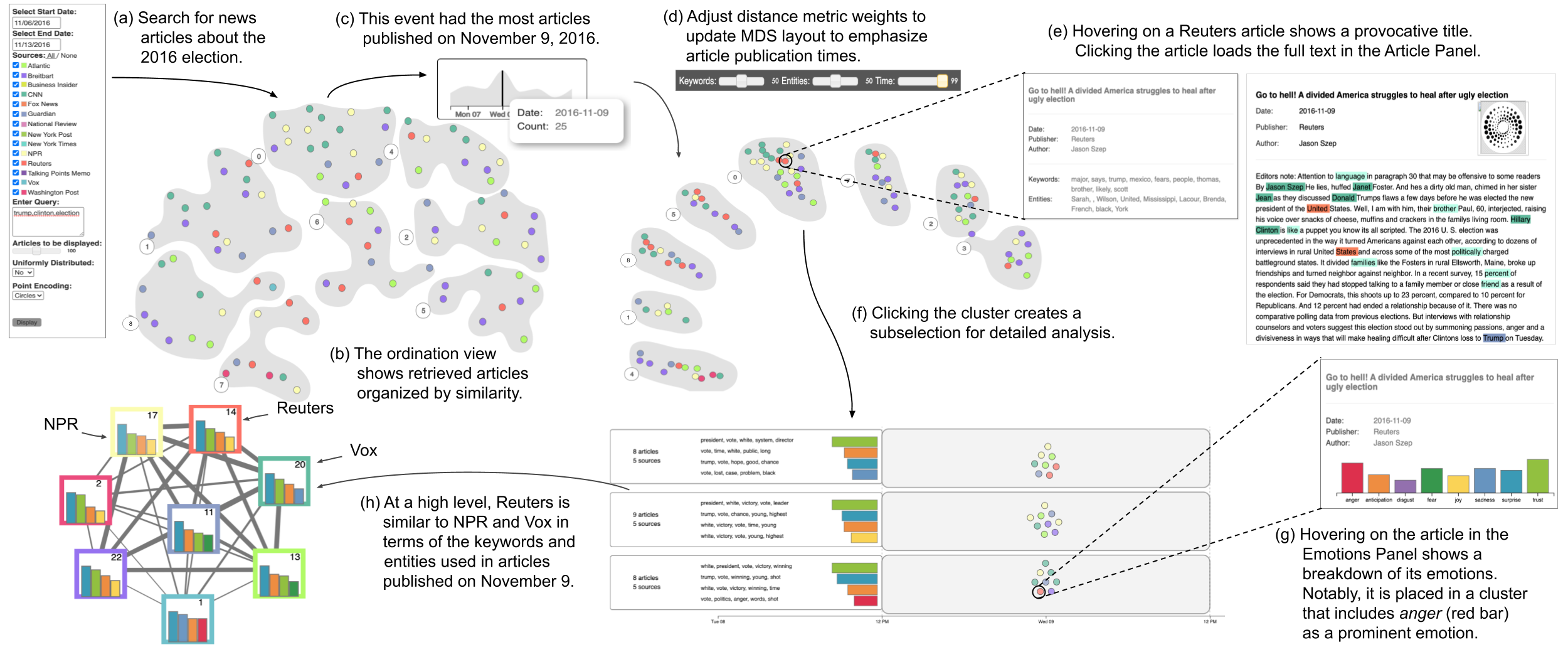}
%   \vspace{-1em}
  \caption{The interactions taken during the Use Case Scenario described in Section~\ref{sec:use_case}.}
  \label{fig:use_case}
\end{figure*}

Two additional overview plots are shown in \textsf{(b6)} and \textsf{(b7)}. These two displays (placed in a tabbed panel) support comparative analysis at the news site-level and between clusters. \textsf{(b6)} The site overview tab shows a graph of all news sites with retrieved articles; edges indicate the overall keyword and entity similarity between the content of their stories. To provide a glimpse of each site's reporting, the top four emotions for each site are plotted in a bar chart (labeled by total stories for each site, with bar colors based on the emotions panel's color palette \textsf{(d2)}). This view provides a summary-level comparison of how news sites differ in their reporting on the current event. The bottom half of this tab shows the most popular keywords and entities for each site.

\textsf{(b7)} The cluster labeling contains a heatmap showing (in rows) the top keywords from each cluster, while (in columns) clusters are columns ordered by index. Cell color indicates the frequency that top keywords appear in each cluster, supporting analysis of high-level coverage trends across the clusters.

% \textsf{(b5)} \aditi{\color{blue}To get a more aggregated view and reduce cognitive load on our users we add two sub panels, Site Overview and Cluster Labels. Site overview provides a graph view of all the sites present in the MDS panel and the pairwise Keyword + Entity similarity between them. Thicker edges represent more similarity in content. And for each site the top 4 emotions are also extracted. The view helps the user get a more summarized view similarities and differences among sites. Cluster label acts like an annotation tool for the shown clusters in the MDS panel. Instead of users reading each and every article of a cluster to understand the cluster, we identify the top 5 most important keywords which best explain it. Thus providing annotations for each cluster. }

% \textsf{(b6)} \aditi{\color{blue} Hovering on number of clusters in the ordination panel shows a line chart of silhouette scores for clusters. Higher the score better is the clustering. This tooltip aids non CS expert users to gain more transparency in the number of clusters and hence leads to a better analysis.}

\textsf{(b8)} Finally, hovering over an article displays its tooltip. To select a subset of articles from the overview (\textbf{DR3}), there are two available interactions. Drawing a lasso makes a freeform selection, while clicking on a cluster selects all of its articles. When an article subselection is made, the NER and emotion panels are populated for detailed analysis.

\textbf{\textsf{(C)} NER Panel.}
The NER panel %\textsf{(C)} 
visualizes the article subselection with \textsf{(c1)} an adjacency matrix and \textsf{(c2)} word cloud. The adjacency matrix shows pairwise article similarities between articles based on the entity distance metric, allowing for direct comparison of two articles. \textsf{(c3)} The user can choose which named entity types are used to compute similarity: persons, locations, organizations, or all \textsf{(c4)}. Entity distance is used for this visualization (instead of keywords) as it bins extracted article text into explicit categories. Below, a word cloud shows which entities are most used in the article subselection. Words are ordered by frequency and colored by entity type. Hovering on a cell in the adjacency matrix highlights shared entities in the word cloud.

\textbf{\textsf{(D)} Emotions Panel.}
The emotions panel %\textsf{(D)} 
visualizes the emotional styles of the article subselection. Articles are clustered via \textit{k}-means clustering using the emotional style vectors. 
% Like the Ordination Panel, the user can adjust the \textit{k}-value. 
\textsf{(d1)} Articles within each cluster are temporally ordered according to their publication day. 
% (i.e., the dominant four emotions based on the averages of articles within that cluster),
% and label the number of articles and news sites that the cluster contains.  
\textsf{(d2)} Using the panel's control widgets, cluster settings can be updated and article circles can be toggled to display as radial bar charts \textsf{(d3)}. This view shows each article's emotional style vector as a set of eight radial bars around its circle.
\textsf{(d4)} Hovering on an article shows this vector using a (standard) bar chart.
\textsf{(d5)} At the left of each cluster, we show that cluster's top-$4$ dominant emotions (as well as the top contributing words to each emotion) to provide a glimpse of the cluster's overall characterization.

% \aditi{\color{blue}Words contributing to the top 4 emotions for each cluster is identified and annotated beside each cluster.}
% the subset of charts selected by the user. The system classifies the selected articles into $K$ user defined clusters. We provide two algorithms, KMeans and Hierarchical for the user to cluster the emotions in. The user can change the number of clusters by a slider with values ranging from 2 to 5. To help users make a faster comparison we also show the individual distribution of emotions present in articles using a radial bar chart. 

% Each row in the emotions panel refers to one cluster with the type of cluster shown beside it. We take the aggregate value of the emotions of all the articles put inside a cluster and show the top 4 emotions dominant in those articles. Hovering on the aggregated emotions shows the name of the emotion and the aggregate percent present in all the articles in a cluster combined. Hovering on the article highlights the same article in the Ordination and the NER panels and also shows a tooltip with the headline and meta data along with a bar chart of individual emotions distribution. 

% \subsubsection{Article Text Panel}
% \label{sec:article_text_panel}
% \vspace{1.5mm}
\textbf{\textsf{(E)} Article Panel.}
The article panel %\textsf{(E)} 
allows inspection of individual articles (\textbf{DR5}) by showing the raw article text and metadata (author, news site, publication date). Articles are loaded in this panel by clicking on an article circle (ordination and emotion panels) or adjacency matrix title (NER panel). To support analysis, the user can highlight the extracted keywords and entities in the article's text.

\vspace{-.5em}
\section{Use Case Scenario}
\label{sec:use_case}

To illustrate how \name{} can be used to analyze coverage diversity, we present a use case scenario with \person, a journalist who is reviewing reporting on the 2016 U.S. presidential election. This event, which took place on November 8, 2016 between Democrat Hillary Clinton and Republican Donald Trump, had reporting both leading up to the day as well as post hoc analysis and reporting on the outcome. Figure~\ref{fig:use_case} shows his workflow.

\textsf{(a)} \person{} first searches for articles from two media organizations (Breitbart and The Atlantic) using the keywords \textit{trump} and \textit{hillary} from November 6--13, 2016.
% and with results limited to the top-100 ranked articles. 
\textsf{(b)} The retrieved articles populate in the ordination panel. 
% \person{} wishes to see how if the 2 sites do differ in terms of the election reporting and in what ways.
\textsf{(c)} By reviewing the temporal distribution of articles, \person{} notices most are published on November 9 (the day after the election). \textsf{(d)} \person{} increases the weight of the temporal distance metric, which updates the MDS layout to place articles published on the same day in the same cluster. 
% \person{} wishes to explore clustered articles of the same day which contains a very dispersed distribution of articles within a cluster of both the news sites. 
\textsf{(e)} Using the tooltip to skim articles in the November 9 cluster, \person{} notices that green circles (Atlantic articles) emphasize the \textit{trump} keyword, however they seem to be downplaying his electoral win (example articles include \textit{Trump's Victory Sends a Disturbing Message About Sexual Assault} and \textit{Empathizing with Trump Voters Right Now}). In contrast, Breitbart articles (in purple) focus on Hillary Clinton's loss (e.g., \textit{Justice has Prevailed with Hillary loss} and \textit{Behind-the-Scenes of Team Trump’s Triumph, Hillary’s Concession}). \textsf{(f)} This is further validated by reviewing the Site Overview Panel; the People entity for The Atlantic focuses on Trump, while Breitbart focuses on Clinton. This suggests that the two sites are exhibiting coverage diversity in how they are covering this event.

\person{} now wants to understand \textit{how} these two sites are differing in their coverage. \textsf{(g)} He uses the Site Overview Panel to review the aggregate emotions expressed by these two sites. Three of the top four emotions (surprise, trust, anticipation) are shared between the sites, however The Atlantic includes fear and Brietbart includes joy as an emotion. \textsf{(h)} To analyze the emotions in depth, \person{} selects the November 9 cluster as an article subselection into the Emotions Panel. He notices an interesting grouping of articles across this panel: the first and second clusters consist primarily of Atlantic articles, and one of the dominant emotions in these clusters is fear (this is supported by the highlighting of keywords in the Article Panel such as \textit{assault}, \textit{war}, \textit{military}, and \textit{revolution}). The third cluster, which has high Breitbart representation, shows joy as a dominant emotion (with keywords such as \textit{victory}, \textit{friend}, and \textit{God}). \textsf{(i)}~Hovering on individual articles further demonstrates this: Atlantic articles tend to have a higher amount of words associated with emotions like anger and fear.

\person{} has now confirmed both that coverage diversity exists in how The Atlantic and Brietbart have covered this event, and he has learned how the polarity has manifested in terms of emotional biases and keywords. The Atlantic has tended towards coverage that evokes anger and fear in the context of Trump's win, while Brietbart shows joy regarding Hillary Clinton's loss.

\vspace{-.5em}
\section{Evaluation}
\label{sec:user_study}

To empirically evaluate \name{}, we conducted a trio of user studies. Study~\#1 consists of pair analytics sessions with $7$ news experts (e1--e7), who are the target demographic for the system. 
% Participants for this study are journalism and political science professors who are experts about reporting and coverage diversity in American news reporting. \aditi{Study~\#1 is specifically made to access the insights given by the journalism experts.} 
% In contrast to running a full usability study, Study~\#2 is streamlined specifically to assess how \name{} insights promoted to journalism experts differ from those for the novices in Study~\#1.
However, to understand if and how \name{} can generalize to a broader population, we subsequently conducted a follow-up study with 11 news novices (p1--p11). In contrast to Study~\#1, we format Study~\#2 as a controlled usability study to understand not only the insights supported by \name{}, but also the overall user experience for users with little background domain knowledge.

% \aditi{To also test the generalizability of the study we perform a second study.} \textbf{Study~\#2} is a controlled usability study with $11$ news novices (p1--p11). It collects several data points to help understand the overall user experience, including system logging, a post-study questionnaire, and in-study commentary in the form of think-aloud comments. In particular, we are interested in learning about what types of insights \name\ might promote for this group. Our assumption is that \name\ effectively supports analysis of coverage diversity, even for users with little background knowledge.

These two studies provided robust insights into how \name{} supports analysis of news events and coverage diversity, but they also highlighted several system features that were either missing or were difficult to understand. After adding a targeted set of new system features based on this feedback, we conducted a third study to validate their utility (Study~\#3).

\vspace{-.5em}
\subsection{Study~\#1 Design}
For Study~\#1, we conducted pair analytics sessions with 7 news experts. Specifically, we wanted to understand what specific types of insights about coverage diversity were supported by \name{}.
% The intent with Study~\#1 was to complement Study~\#2 by assessing if journalism expertise resulted in new types of insights or usage patterns promoted by \name{}.
% Specifically, Study~\#1 found that participants could recognize coverage diversity, but had trouble connecting it to strong conclusions about bias and framing.
% \footnote{While an in-person, think-aloud study that mirrored Study~\#1 was originally intended, due to COVID-19 concerns this was not possible.} 
Pair analytics~\cite{arias2011pair, elmqvist2015patterns} is a procedure for capturing reasoning processes in visual analytics. A visualization expert well-versed with an interface's functionality ``drives,'' while the study participant reasons and makes decisions based on their domain expertise. Verbal discussion between the driver and participant is the basis for understanding of the participant's sensemaking process as well as what specific insights are uncovered during investigation.

\textbf{Study Procedure.}
Study sessions were conducted remotely using Zoom videoconferencing software. Kaleidoscope's browser window was shared via Zoom's sharing feature from the visualization driver's computer to the remote participant. QuickTime Player recorded audio discussion and screen capture.

For each session, the visualization expert first gave an overview of the system's functionality until the participant felt confident to proceed. After this, the news expert could freely analyze and explore the system using the entirety of the \textit{All the News} text corpus. Sessions lasted as long as the news expert desired, though it was suggested they spend at least 15 minutes in their investigation. At the end of each session, participants were asked to provide freeform commentary and feedback on the system, such as what features they liked, disliked, and found useful.

\textbf{Participants.}
The 7 participants in Study~\#1 were professors at Arizona State University in the political science and journalism departments (5 assistant and 2 associate professors). All participants were highly knowledgeable about American news reporting, including coverage diversity and perceived biases of news sites; a couple of the faculty specifically researched media bias and coverage diversity in reporting.

\vspace{-.5em}
\subsection{Study~\#1 Results}
\label{sec:study_1_results}

To understand how \name{} promotes insights to news experts, we reviewed the discussions from recorded screencasts. We report here the usage patterns and the insight themes that were promoted.

\textbf{\name{} supports scalable analysis of news events.}
An immediate recognized benefit, noted by all participants, was that \name{} quickly allows them to see and analyze a large number of stories about a news event:
``\textit{The fact that there are so many articles and site at the same place makes it useful already}'' (e1).
``\textit{The system would be a good place to start for journalists because we deal with a lot of news articles and it would definitely be easier if such a system would be used}'' (e2).
``\textit{Overall I would use this system to start with a big dataset analysis because it seems really good for that}'' (e5).
``\textit{This system would definitely be a great place to start. So many articles and news sites in one place. I find this system really comprehensive}'' (e3).

Multiple participants quickly realized that articles could be ingested at by reviewing keywords instead of reading the entire text: ``\textit{Visualizing keywords and entities like this is really great. I don’t have to read the entire article at all}'' (e4). ``\textit{The keywords and entities are a good thing in the system, it very nicely shows me which article is similar and which is it different from}'' (e1).

Several participants mentioned that \name{} showed coverage diversity through the lens of \textit{polarity} between news sites and over time. ``\textit{Bias isn't explicitly shown, but as a user walks through or uses the system more I think it would be easier to see polarities using the emotion panel \ldots\ the system gives me a very good idea of coverage}'' (e1).
``\textit{This would help me a lot in the kind of work I am doing. I work on the usage of words in a news article to then find the underlying polarity \ldots\ this system would be just apt for it}'' (e4). 

\textbf{The visual analytics were a novel experience.}
While one participant (e6) had previous experience using LDA~\cite{blei2003latent}, no participants had used a visual analytics system similar to this as a part of their own work or research. ``\textit{I have never used such a system before, I have done topic modeling, but this is interesting and new}'' (e6). ``\textit{I think just the fact that so many features and site comparison exists in the system makes it interesting because we generally don’t have anything like this to use}'' (e2). ``\textit{I really like the emotions panel. It’s a very new take and it shows me a broad spectrum over time, which is really informative for me and my research}'' (e5).
% While (e6) noted ``\textit{I probably wouldn't use [the system] for my work,'' he suggested that ``maybe a political psychologist would find the system more useful because of the emotion clusters.}''

% Additionally, the news experts leveraged their knowledge of coverage diversity to nuance about how the system decomposed articles into keywords, entities, and emotions. 

\textbf{An unexpected usage of \name{} as a validation system.}
While data exploration and discovery are common visualization goals, empirical \textit{validation}---such as of a hypothesis or model's performance---is also a function that visualization can support. Since news experts had existing knowledge and intuitions of coverage diversity (as well as the potential biases of news sites), several participants felt the system could be used as a verification tool. 
% Four participants explicitly said that they could use the system for validation in the context of their existing knowledge.
``\textit{Since I already know the bias or the kind of articles sites publish, I might use this system to compare them and validate my understanding}'' (e2). Interestingly, e7 mentioned using the system as a way to review news articles that she helps create: ``\textit{We can use it to understand what’s going on: model a [news] outlet. It might be helpful for comparison. How are we doing in presenting news? Are we similar to CNN or NPR?}''

\textbf{Targeted suggestions by news experts.} While participants could understand the visualizations and interactions, some took a bit of time to acclimate to the workflow and the system's available functionalities. ``\textit{Adding a few breadcrumbs might help, because they are so many things going on here}'' (e5). Likewise, as non-engineers, participants were sometimes unsure of the backend algorithms. For example, clustering was positively reviewed, but several participants (e1, e2, e4, e5) suggested adding some form of explanation as to how the metrics were computed; e.g., why the specific number of clusters was chosen: ``\textit{A little extra support for people with no CS background would help, because its a very new thing for me}'' (e4).  ``\textit{A lot of non-computer scientist people might find it a little overwhelming}'' (e3).
% Like the news novices in Study~\#1, all news experts in this study were also visualization lay users, and some were initially overwhelmed by the amount of features and interactions, especially as non-engineers. 
% In contrast to Study~\#2, news experts did not suggest explicitly labeling news sites or articles by perceived biases. 
Another feature that was requested multiple times was summary and comparative views, to show aggregate event behavior and site-level comparisons:
% That said, several suggestions were given in the context of assisting their research and own analysis workflows for investigating article collections. One was adding additional comparative visualizations to directly ``\textit{focus on the differences among articles}'' (e3). Likewise, summary visualizations were also requested to show aggregate event behavior and site-level comparisons. 
``\textit{`I preferred more overview \ldots\ [Start by showing] just showing one cluster first and then the user digs in deeper}'' (e5). ``\textit{I wished it directly showed me the sites' differences or similarities}'' (e3).

\vspace{-.5em}
\subsection{Study~\#2 Design}

Study~\#1 provided understanding into how \name{} can be used by domain experts to analyze coverage diversity. However, during development we also wondered if such a system can benefit news novices---users that are largely agnostic about the coverage diversity of specific news sites. As a motivation for this, a recent RAND report indicates that people are increasingly searching for ``alternative views'' about newsworthy events of interest~\cite{pollard2019profiles}.  Therefore, we conducted a second study targeting these news novice users. 

In contrast to Study~\#1, and to better regulate the user experience and understand the thought process of news novices, Study~\#2 was run as a full qualitative study that included both in-study think-aloud protocol and a post-study questionnaire and feedback session about the user experience
% , and logged system interactions. 
% Due to space constraints, the system logging and think-aloud analysis can be found in the supplemental materials, and 
This allows us to understand both how the insights of novice users compare and contrast to domain experts, and also to understand the overall system experience for a set of domain-agnostic users.

To begin Study~\#2, participants first completed a survey to collect demographic information and background knowledge. After this, the following three stages were run:

\textbf{Training Stage.}
Each participant was given a hands-on tutorial describing all features and interactions in \name. 
% A proctor explained its features and interactions. 
Participants could then practice with the system for as long as desired, analyzing news events in the \textit{All the News} corpus that occurred in January 2017. A list of four noteworthy events was provided to the participant as a reference point. The intent was for the participant to become comfortable using the interface before proceeding to the main study.

\textbf{Exploration Stage.}
Next, each participant was free to explore the news corpus using \name{} for as long as they desired. It was suggested that participants spend at least 15 minutes using the system, however there was no hard cutoff. To  constrain the study design, we restricted article queries to news events that happened in the latter half of 2016 (approximately 62,700 total articles). Participants were instructed to start by exploring and analyzing the coverage of two significant news events from that year: (1) The 2016 presidential election on November 8. (2) The Pulse Nightclub shooting in Orlando, Florida on June 12. After, participants were free to explore and analyze any news events they wanted. A list of noteworthy news events (with associated keywords) was provided for reference, though participants could query and use the system however they desired. During this stage, participants used think aloud protocol to verbalize their thought processes and actions~\cite{fonteyn1993description}.

% Notable news events from this period include  the Brexit vote (June 23), the Nice truck attack in France (July 14), the 2016 Summer Olympics in Rio de Janeiro, Brazil (August 5--21), and the Chicago Cubs winning the World Series (November 2). 
% which notably is the year of the U.S. presidential election between Donald Trump and Hillary Clinton, which was subject to intense political coverage.

\textbf{Review Stage.} Finally, participants completed a short survey questionnaire about system impressions and functionality using a Likert scale (1 – strongly disagree, 7 – strongly agree). Participants then had the opportunity to provide freeform comments, suggestions, and criticisms about the interface and their experience.

\begin{figure}[t]
  \centering
  \includegraphics[width=.8\columnwidth]{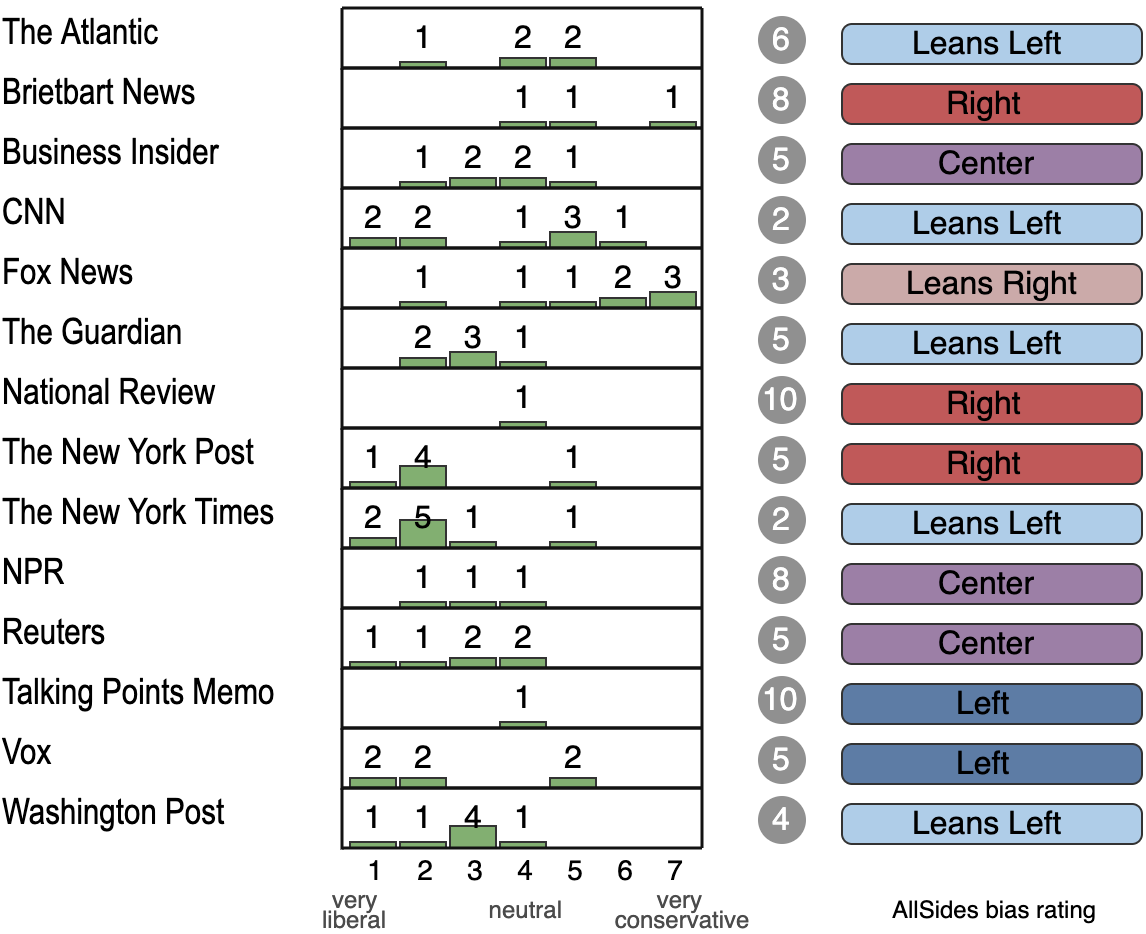}
  \caption{
    Before taking Study~\#2, novice news participants were asked to rate the perceived political biases of the news sites in the study's data corpus. The histograms show the overall ratings for each site; gray circles indicate the number of participants who simply said ``I don't know.'' This figure illustrates that Study~\#1 participants were largely unfamiliar with American news sites, and could therefore be considered novice news users. As a reference, the rightmost column shows each site's media bias rating from AllSides~\cite{allsides}.
  }
  \label{fig:study_1_prestudy_responses}
\end{figure}

\begin{figure}[t]
  \centering
  \includegraphics[width=.85\columnwidth]{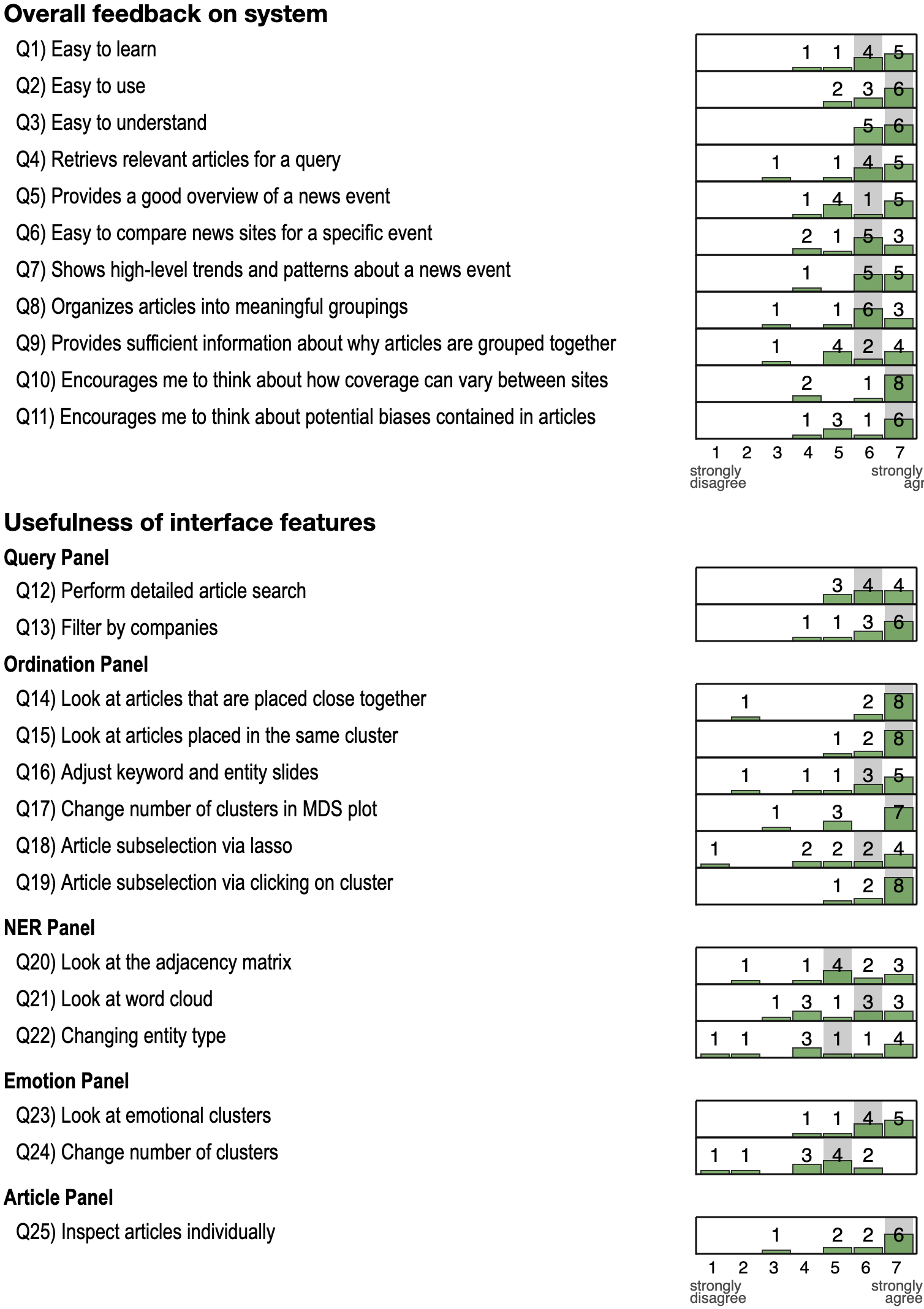}
  \caption{
    This chart shows overall ratings from the post-study questionnaire about various system aspects by the novice users in Study~\#2. Median ratings are indicated in grey.
  }
  \label{fig:survey_ratings}
\end{figure}

\textbf{Participant Recruitment and Apparatus.}
We recruited $11$ non-domestic engineering students from $<$\textsf{Anonymous University}$>$: age $\mu = 24.27$ years ($\sigma = 1.95$), 7 males and 4 females, most of whom (8/11) had only moved to the United States after 2016. All participants were proficient in English and had good (corrected if necessary) eyesight. 
% Based on a 7-point Likert scale, participants self-reported high comfort in 
% doing data analysis ($\mu = 5.5$, $\sigma = 1.81$) and 
% reading basic charts and visualizations ($\mu = 6.01$, $\sigma = 1.55$). 
\name\ was displayed in Google Chrome 
% (full screen mode) 
on a $24$-inch monitor ($3840\times2160$ resolution) with keyboard and mouse and connected to a MacBook Pro running macOS Mojave. QuickTime Player recorded session screencasts and audio.

To determine if Study~\#2 participants could be considered news novices, we included a pre-screening as a part of the background knowledge collection.
As all participants were non-domenstic students, this included asking several questions about American news reporting. While many participants reported some familiarity with American news companies based on a 7-point Likert scale ($\mu = 4.58$, $\sigma = 1.53$) and 8/11 participants agreed that they regularly kept up with American news stories, when we asked participants questions about specific news sites they were largely oblivious about how to rate news sites.  Figure~\ref{fig:study_1_prestudy_responses} shows this explicitly; the high variance in the historgrams indicate uncertainty, and the gray circles to the right indicates the number of participants who simply answered ``I don't know'' about a particular site. The labels in the right-most column indicate the site's AllSides media watchdog rating, which likely represents the consensus rating for news experts. While bias is not a perfect proxy for the coverage diversity analysis that \name{} supports, this chart effectively illustrates that the Study~\#2 participants were largely unfamiliar with the American news sites in our data corpus, and could therefore be considered news novices for the purposes of the study.

% there was high uncertainty and abstaining when participants were asked to rate the study's news sites on a spectrum of left-to-right 
%perceived 
% political leanings.  shows these results, which indicate a general lack of knowledge about perceived biases and coverage diversity within American news companies.
% and no well-defined frame of reference to contextualize news site-specific coverage diversity.

\vspace{-.5em}
\subsection{Study~\#2 Results}
\label{sec:study_1_results}

To analyze Study~\#2, we first briefly report high-level system ratings based on post-study questionnaire responses, and then analyze participant comments (collected both from think-aloud comments and post-study feedback) to qualitatively assess the types of insights and action patterns that \name{} promotes for novice users.

\subsubsection{System Ratings via Questionnaire Responses}
\label{sec:study_2_survey_results}

Figure~\ref{fig:survey_ratings} shows questionnaire responses about \name{} which describe overall system feedback and the perceived usefulness of its interface features. Overall system ratings (\textsf{Q1}--\textsf{Q11}) were generally positive, including that it was easy to learn, use, and comprehend (\textsf{Q1}--\textsf{Q3}), organized data into a high-level overview (\textsf{Q5}), supported meaningful analysis (\textsf{Q6}--\textsf{Q9}), and encouraged participants to think about coverage diversity in news reporting (\textsf{Q10}--\textsf{Q11}).

Responses about specific interfaces features (\textsf{Q12}--\textsf{Q25}) followed a similar trend in being generally well regarded. However, reviewing these responses individually indicates at least some features were not as well received, notably the word cloud in the NER panel (\textsf{Q21}--\textsf{Q22}) and the action of changing the number of clusters in the emotions panel (\textsf{Q24}).

\subsubsection{Supporting Insights for Novice Users}
\label{sec:study_2_insights}

To understand the insights of news novices, we reviewed both the think-aloud comments made during the exploration stage and the freeform comments from the review stage. Specifically, we contextualize these results to Study~\#2 to highlight the different experiences that \name{} provides for non-domain experts.

\textbf{For news novices, the interface was both easy to use and too complex.}
Despite not being the intended user base, we received several comments about the \name{} being 
% ``\textit{cool}'' (p11), 
``\textit{useful}'' (p6), and ``\textit{easy to use}'' (p7). However, some users paradoxically felt overwhelmed by the system's available features and interactions. ``\textit{I didn't use all the features as there were so many}'' (p11). Multiple users suggested streamlining the user experience (p7, p6, p1), or thought that follow-up sessions would lead to more efficient analysis: ``\textit{Probably [the] user needs more time to make full use of the system}'' (p5). This mixed feedback is likely due to a lack of domain knowledge by news novices, which hampered some of the users' experiences while not being a problem for others.

\textbf{Aspects of the user experience echoed news experts.}
Similar to Study~\#1, several participants remarked how \name{} allows them to review articles in a scalable fashion:``\textit{I can generally just read one article from one site on my news app or site, however this system is helps me read a number of articles on an event by different sites at the same time. This in addition to the other features such as keywords, time, entities and emotions is really helpful}'' (p8).

Entities, both via tooltips and in the NER panel, were especially helpful for getting a quick overview of a story: ``\textit{Entities tell me who and what all are included without reading the entire article}'' (p6). Some users leveraged the entity-based pairwise similarity in the NER panel's adjacency matrix, as this was the only visualization that directly compared two articles: ``\textit{There is more similarity in news articles on people than other two}'' (p3). 

% One useful feature was the timeline widget in the Ordination Panel, which several users leveraged for temporal analysis and introspection about event reporting. ``\textit{Even though Trump became president on 8th, most articles were covered on the on 16th. Not sure why}'' (p3). ``\textit{The timeline density suggests that as the incident was explored more more articles came up, than at the time of the articles}'' (p4).

\textbf{Clustering is effective for showing coverage diversity.}
Several participants described the clustering---both in the ordination and emotions panels---as an effective and ``\textit{intuitive}'' (p7) way to show coverage diversity. ``\textit{When I look at the clusters [in the MDS plot], there is a shift in polarity from right to left. Here towards the right there are more articles on Hillary and here’s there’s more Trump}'' (p11). In the MDS plot, clustering ``\textit{helped to select which articles to read}'' (p8) and make subselections on. 

In the emotions panel, clustering ``\textit{allowed me to get a good idea of emotional values from the news articles}'' (p6). Several participants mentioned the emotional style vectors provided helpful insight into how and why articles were clustered, thus making the emotions panel ``\textit{better than clustering based on keywords and entities as it was more intuitive than keyword clustering}'' (p6). 

% In general, clustering by computed emotional styles was useful to interpreting news articles and understanding coverage diversity. ``\textit{The cluster points tell me a lot about the diversity of the sites}'' (p11). ``\textit{Emotions are very much attuned with the articles}'' (p2). ``\textit{The tone of the article makes sense with the emotional cluster}'' (p6). 
At times however, there was confusion when computed clusters did not line up with participant expectations or mental models. For example, in the ordination panel participants would sometimes wonder why a particular story was included in a particular cluster, and in the emotions panel participants sometimes were unsure how emotions contributed to a cluster. As an example, when p2 was reviewing articles about the Pulse Nightclub shooting: ``\textit{It's all fear and anger and surprisingly there is no sadness which one would normally expect. I didn't expect that.}''

\textbf{Participants recognized diversity, but it was hard to translate these into broader conclusions about media bias.}
Along these lines, while participants were able to recognize variation between articles and sites, they were reluctant to use these to draw explicit conclusions about media bias. As most users were unfamiliar about the perceived political leanings of media sites, many were unwilling to make strong statements about specific sites or bias based on keyword, entity, and emotional variations in the stories.
% Regarding the 2016 presidential election, (p5) tepidly reported ``\textit{Maybe Brietbart supports Trump}'' (p5). Instead, 
Statements tended to focus on the specific articles that were being analyzed: ``\textit{Not sure if CNN is anti- or pro-[Trump], but its articles show that they are doubting Trump's win}'' (p7).

Interestingly, some participants contextualized news sites using a prior they were familiar with: clickbait writing. ``\textit{Atlantic has clickbait headlines}'' (p2). ``\textit{CNN kind of publishes reactionary kind of headlines but Reuters has a more balanced view}'' (p4). ``\textit{This site has clickbaity articles, the headlines would make you think something but the emotions in the tooltip suggest something else}'' (p2). Several participants also suggested the system explicitly anchor or label the bias and framing of articles or news sites to help contextualize how they should be interpreted. ``\textit{Naming the clusters would have made user investigation faster}'' (p6). ``\textit{I see how on an event is covered, but I couldn't form an opinion [about specific news sites] \ldots\ so if more annotations were present it would have been better}'' (p5).
% We discuss potential implication of this in Section~\ref{sec:discussion}.

% Along these lines, some styling features were explicitly disdained. Several users tested out encoding articles using their site logos, but disliked them compared to the default colored circles: ``\textit{I like the circles better than the logos. The visual encoding is easy to understand}'' (p3).

\vspace{-.5em}
\subsection{Study~\#3}
Studies~\#1 and \#2 validated several aspects of \name{}, particularly its ability to support coverage diversity for news experts (as well as its ability to introduce the topic to news novices). Despite the largely positive feedback, some features were commented upon as either missing or confusing by multiple study participants. Specifically, we identified three targeted system improvements based on participant feedback: (1) provide additional transparency or explanation into the number of clusters to show, both in the ordination and emotions panels, (2) provide views that support summary and comparative analysis of news events and sites, and (3) provide additional explanation of how emotions are computed. The systems features in Figure~\ref{fig:interface}\textsf{(b5, b6, b7, d5)} were thus added to the system.

To validate these design improvements, we followed up with four participants (u1--u4) from Studies~\#1 and \#2 (three from Study~\#1, one from Study~\#2), who could compare the system's updates against the original version. For each participant, we conducted a pair analytics session over Zoom, demoing the system's new functionality and soliciting freeform feedback. In general, the new components were positively appreciated. ``\textit{I like the site overview part because, the first time I used the system, it felt like a lot of information was thrown at me and I didn't know where to start from}'' (u1). ``\textit{The cluster annotation is really good. Earlier it was a little harder than the present system}'' (u3). All three expert users (u1--u3) reiterated their earlier comments that systems like \name{} would be useful for their work, highlighting the lack of accessible visual analytics systems for studying news corpuses: ``\textit{I still continue to think this will be a useful tool. It's something I could see myself using in research.}''(u3). 

\vspace{-.5em}
\section{Discussion}
\label{sec:discussion}

While there are many existing text visualization tools in existence, based on our experience in developing and evaluating \name{}, we believe that interfaces tailored specifically for news reporting analysis can better benefit both for news experts and news novices. \name{} design was intended to elicit nuanced analysis of coverage diversity via the subselection workflow. To understand the implications of this design study, we comparing the insights and workflow themes between the news expert and novice users, and discuss how they can guide the design of visualization systems for news reporting analysis.

\textbf{Novice news users recognize diversity, news experts can see research possibilities.}
While both user groups agreed that \name{} supports the analysis of coverage diversity, one main takeaway from the studies is that insights differ based on the user's experience. Novice news users were able to assert that coverage diversity exists in the articles written about news events, but generally could not make strong conclusions or broad inferences about bias. News experts, already familiar with coverage diversity and media bias, were able to use the system to see deeper nuances about coverage diversity. Analysis systems for news novices can be tailored to accommodate their lack of domain expertise.
% For some participants in Study~\#2, \name{} was likely too advanced, while news experts were able to leverage their prior knowledge in coverage diversity to help contextualize their analyses.
% In this way, extracted keywords, entities, and emotions did not end up being perfect proxies for bias, but they paired well with domain knowledge to facilitate in-depth analysis and for potential research. 

\textbf{Explainability on demand improves sensemaking.}
No users in our evaluation were experts in data visualization or NLP. While both user groups could understand and interact with the interface's visualizations, many participants wanted more explainability, particularly when data points or clusterings displayed in unintuitive or unexpected ways. Such desires echo recent trends in explainable machine learning and artificial intelligence, where transparent and interpretable models are desired to promote trust in and understanding of predictions~\cite{lipton2018mythos}. The explainability features added for Study~\#3 were non-disruptive to the overall user experience, but they were positively received in our follow-up sessions with participants. 
% mittelstadt et al note that to make models more trustworthy and accountable, explanations should communicate why a model is making certain predictions in ways that `everyday users' (ie, data lay users) can understand.

% \textbf{Contextualizing \name{} by task abstractions and visualization lay users.}
% \name{}'s design was based on a task abstraction for visualization lay users. As such, it employs techniques meant to be accessible to the general public. Based on our study results, all of tasks \#1--6 were effectively supported. At a high level, these tasks enable the investigation and analysis of coverage diversity. While this goal was supported for both novices and experts, each group developed their own (unexpected) spin on how to user the system. News novices contextualized coverage diversity by recognizing clickbait reporting. News experts saw the system as a way to validate their existing notions and datasets.

% Regarding the use of \name{} by visualization lay users, the system received mixed feedback about complexity, as it was paradoxically both comprehensive and overwhelming depending on the user (though such users felt that follow-up sessions with more acclimatization would overcome this issue). Study~\#2 especially provides an interesting takeaway. 
\textbf{The need for visualization tools in journalism research.}
Study~\#1 participants were researchers in journalism and political science, but none had previously used visual analytics tools like this. While systems similar to \name{} have previously been published in the visualization community, the lack of adoption by our participants indicates that wider dissemination would benefit other communities.  Tools like this also motivate an interesting question for these communities: \textit{If coverage diversity is identified, how can  we reduce media bias?} This is a non-trivial problem, but we believe visualization can likely provide an important step in the process, by helping researchers identify where issues are present.

\textbf{Visually analyzing the temporal dynamics of news coverage.}
Some participants in Study~\#1 noted temporal analysis as an important facet of coverage diversity: ``\textit{Development of news stories over time is interesting \ldots I personally for my research loved the time evolution in both the cases especially the emotions over time, that is a really informative information in journalism}'' (e7). While \name{} provides a limited amount of temporal analysis, systems that focus on the temporal (and unique) dynamics of news stories are lacking. For example, reporting about events evolves over time as more facts are learned and media analysis and commentary is conducted. Future visual analytics systems can be tailored towards this type of analysis, and we intend to extend \name{} to better focus on this type of evolutionary analysis.

% with the implication that systems that provide robust analysis in this domain would be beneficial. 
% Finally, temporal analysis was noted as important facet of coverage diversity, with the implication that it could additionally be expanded or emphasized. ``\textit{Development of news stories over time is interesting \ldots I personally for my research loved the time evolution in both the cases especially the emotions over time, that is a really informative information in journalism}'' (e7). We discuss this type of visual analysis in Section~\ref{sec:discussion}.

% While there are many visual analytics systems that support temporal analysis and document corpus analysis, to our knowledge there is little that supports the specifics of news coverage and some of its unique data semantics. 
% For example, in contrast to general document datasets, such as a collection of books, news articles generally orient around an event of interest, and therefore subsets of articles can be clustered and analyzed together based on their reporting target. How the articles report on the event (the coverage diversity) is a much more specialized analysis than general text/document analysis. In particular, reporting about events evolves over time as more facts are learned and media analysis and commentary conducted. Future visual analytics systems can be tailored towards this type of analysis, particularly for news experts who research media and journalism.

\textbf{Explicitly labeling bias in news stories?}
Several news novices requested explicit bias labeling of articles and news sites, however we are cautious that such an approach is necessary or even appropriate. \textit{Ex ante} labeling inherently reduces the power of empirical analysis by anchoring users to the classification outputs of a model.
% (In fact, such was pointed out by study participants regarding $k$-means values and emotion clusters, which lead to the new explainability features for Study~\#3.) 
When considering coverage diversity in the context of specific news events, explicit bias labeling might poorly account for sites that publish contrasting viewpoints such as editorials and opinion articles. Interfaces that can provide an initial bias labeling while accounting for such nuances (perhaps via uncertainty or probabilistic techniques) are one strategy, but we leave this as future work.

\textbf{Alternate visual encodings for \name{}} Though we found that \name{} successfully supports the analysis of the coverage diversity of news articles, in the future we intend to look at ways to improve the analytic process via alternative visualization and analytical approaches. As an example, instead of showing a flat clustering of retrieved articles (in the Ordination Panel), a hierarchical topics tree could instead be constructed, with individual nodes representing topics constructed from relevant keywords, entities, and/or biases. Coverage diversity could be demonstrated by analyzing and interacting with the tree's structure; for large trees, aggregation and simplification techniques would likely need to be employed (e.g.,~\cite{wongsuphasawat2017visualizing}).

\vspace{-.5em}
\section{Conclusion}

We contribute \name{}, an interactive visual analytics system that supports analysis of news events with a focus on the coverage diversity of reporting articles. \name{} is designed based on a formal task abstraction for news experts and combines several NLP and visualization techniques into an accessible user experience based around keyword subselections in article sets. Based on a holistic, three-part evaluation, we find that \name{} supports different types of insights based on the journalistic domain expertise of the user: news novices are able to recognize coverage diversity at a high-level, but news experts can contextualize it for deeper insight, such as by characterizing articles according to their extracted emotional styles. Study results motivate several guidelines and takeaways for future systems that visualize coverage diversity, including the importance of explainability for non-technical users and that the temporal, comparative, and summary dynamics of news events and coverage can be emphasized.

\bibliographystyle{abbrv-doi}

\bibliography{template.bib}
\end{document}